\documentstyle[12pt,aaspp4,flushrt]{article}
\begin{document}

\title{Determining the Amplitude of Mass Fluctuations in the Universe}

\author{Xiaohui Fan, Neta A. Bahcall, and Renyue Cen\\
Princeton University Observatory\\
Peyton Hall\\Princeton, NJ
08544-1001\\
electronic mail : fan,neta,cen@astro.princeton.edu}

\section*{Abstract}
We present a method for determining the {\em rms} mass
fluctuations on 8 $\rm h^{-1}Mpc$ scale, $\sigma_{8}$.
The method utilizes the rate of evolution of the abundance of rich clusters
of galaxies.
Using the Press-Schechter approximation, we show that the
cluster abundance evolution is  a strong function of $\sigma_{8}$: $d\log n/dz \propto
-1/\sigma_{8}^2$; 
low $\sigma_{8}$ models evolve exponentially faster than high $\sigma_{8}$ models,
for a given mass cluster.
For example, the number density of Coma-like clusters decreases by a factor
of $\sim 10^{3}$ from $z = 0$ to $z \simeq 0.5$ for $\sigma_{8}$=0.5 models,
while the decrease is only a factor of $\sim 5$ for
$\sigma_{8} \simeq 1$. 
The strong exponential dependence on $\sigma_{8}$ arises because  
clusters represent rarer density peaks in low $\sigma_{8}$ models.
We show that the evolution rate at $z \lesssim 1$ is insensitive to
the density parameter $\Omega$ or to the exact shape of the power spectrum.
Cluster evolution  therefore provides a powerful constraint on $\sigma_{8}$.
Using available cluster data to $z \sim 0.8$, we find  $\sigma_{8} = 0.83 \pm 0.15$.
This amplitude implies a bias parameter
 $b \simeq \sigma_{8}^{-1} = 1.2 \pm 0.2$,
i.e., a nearly unbiased universe with mass approximately tracing light on large scales.\\ \\
{\em subject headings :} galaxies : clusters -- galaxies : evolution --
galaxies : formation -- cosmology : theory -- 
 large scale structure of universe
\section{Introduction}
Density fluctuations in the early universe provide the
initial seeds for  the structures we see today.
Various observations of large-scale structure constrain the {\em shape} of the 
power spectrum of fluctuations on different scales 
(Fisher {\em et al.} 1993, Park {\em et al.} 1994, Peacock \& Dodds 1994, 
Strauss \& Willick 1996, Landy {\em et al.} 1996),
but the {\em amplitude} of the mass fluctuation spectrum 
has been difficult to obtain. (The microwave background anisotropy 
provides so far the only direct amplitude measure on large scales; Smoot {\em et al.}
1992, Bunn \& White 1997).
Knowledge of both the shape and amplitude of the spectrum is needed
before critical cosmological implications can be derived.

 The amplitude of the mass fluctuations provides an important constraint on the
relative distribution of mass and light in the universe, 
which remains one of the central unresolved problems
in cosmology: does mass follow light on large scales (an unbiased universe) or
is mass distributed differently than light (biased universe)?
The bias parameter (Kaiser 1984) is quantified as $b= \delta_{gal}/\delta_{m}
\simeq \sigma_{8}(gal)/\sigma_{8} \simeq 1/\sigma_{8}$, 
where $\delta = \Delta\rho/\overline{\rho}$ is the overdensity (in galaxies or mass, 
respectively), $\sigma_{8}$ is the {\em rms} mass fluctuation 
within a top-hat radius of  $8 \rm h^{-1}Mpc$,
and  $\sigma_{8}(gal) \simeq 1$ is the observed optical  galaxy
fluctuations on that scale (Davis \& Peebles 1983). 
A bias of $b \simeq \sigma_{8} \simeq 1$ implies an unbiased universe,
where mass follows light on these scales (assuming a scale-independent bias on
large scales), 
while $b \simeq 2$ ($\sigma_{8} \simeq 0.5$) implies a highly
biased universe, with mass distributed more widely than light.

Galaxy clusters are useful tools in  
cosmology. Since clusters represent high density peaks, 
their present-day abundance directly reflects the amplitude of density fluctuations on the relevant cluster mass scale;
the latter corresponds to the mass enclosed within 
$R \sim 8 \rm h^{-1}Mpc$ spheres in the early universe.
Observations of the present-day cluster abundance indeed place one of the
strongest constraints on cosmology and the amplitude of mass 
fluctuations : $\sigma_{8}\Omega^{0.5} \simeq 0.5$ (Bahcall \& Cen 1992,
White {\em et al.} 1993, Viana \& Liddle 1996, Eke {\em et al.} 1996, Pen 1997).
This relation, while powerful, is degenerate in $\Omega$ and $\sigma_{8}$;
neither parameter can be directly determined from this constraint.
Models with $\Omega=1$ and $\sigma_{8} \simeq 0.5$
are indistinguishable from models with
$\Omega \simeq 0.25$ and $\sigma_{8} \simeq 1$. 

In the present paper we investigate a method that can measure, for the
first time, the amplitude $\sigma_{8}$, nearly independent of $\Omega$.
The method is based on the evolution rate of cluster abundance to $z \sim 1$;
this evolution breaks the degeneracy that generally exists between $\Omega$ and
$\sigma_{8}$ and allows the first determination of each of the parameters 
independently (see also Press \& Schechter 1974, 
Peebles {\em et al.} 1989, Henry {\em et al.} 1992, Eke {\em et al.} 1996,
Oukbir \& Blanchard 1997, Carlberg {\em et al.} 1997, Bahcall, Fan \& Cen, 1997).
In our previous paper (Bahcall {\em et al.} 1997) we presented results from 
cosmological simulations. 
In the current paper, we develop the theoretical basis that shows
 that the evolution rate of cluster abundance is strongly dependent on
$\sigma_{8}$ and  insensitive to $\Omega$ or to the exact shape of the
power spectrum (to $z \lesssim 1$). Therefore $\sigma_{8}$ can be 
determined by observations of cluster evolution.
We present the method in \S 2, and  determine $\sigma_{8}$ in \S 3.
\section{Dependence of Cluster Evolution on Cosmological Parameters}
The Press-Schechter (1974) formalism approximates the evolution of the comoving
number density of halos with mass M ({\em mass function}) in an initial
Gaussian density field assuming hierarchical structure formation from 
linear perturbations:

\begin{equation}
\frac{dn}{dM} = \left(\frac{2}{\pi}\right)^{\frac{1}{2}} \frac{\overline{\rho}}{M^{2}} \frac{\delta_{c}}{\sigma(z,M)}\left|\frac{d\ln\sigma(z,M)}{d\ln M}\right|
\exp\left(\frac{-\delta_{c}^{2}}{2\sigma(z,M)^{2}}\right),
\end{equation}
 where $\sigma(z,M)$ is the linear theory {\em rms} mass density fluctuation in
spheres of mass $M$ at redshift $z, 
 \delta_{c} \simeq 1.68$ is the critical density contrast needed for collapse
(only weakly dependent on $\Omega$,
Eke {\em et al.} 1996, Kitayama \& Suto 1996), and $\overline{\rho}$ is the
mean cosmic density. 
The mass refers to the virial mass of the system.
Here
\begin{equation}
\sigma(z,M) = \frac{\sigma_{0}(M) D(z,\Omega,\Lambda)}{D(z=0,\Omega,\Lambda)},
\end{equation}
where $D$ is the linear growth factor (Peebles 1980) that  represents
the growth of a perturbation in the linear regime. For $\Omega = 1$,
$D \propto (1+z)^{-1}$; $D$ grows slower for $\Omega < 1$. 
$\sigma_{0}(M)$ is the present {\em rms} mass fluctuation within sphere of mass $M$, 
\begin{equation}
\sigma^{2}_{0}(M) = \frac{1}{2\pi^{2}} \int dk k^{2} P(k) |W_{R}(k)|^{2},
\end{equation}
where $P(k)$ is the mass density power spectrum and $W_{R}(k)$ is the window function. If $P(k)$ is
 approximated as a power law $P(k) \propto k^{n}$ around 8 $\rm h^{-1}$ Mpc scale then
\begin{equation}
\sigma_{0}(M) = \sigma_{8} \left(\frac{M}{M_{8}}\right)^{-\alpha} \propto
\hspace{3mm} \sigma_{8}M^{-\alpha}\Omega^{\alpha},
\end{equation}
where $\alpha=(n+3)/6$, $M_{8}$ is the mean mass within a sphere of radius $8 \rm h^{-1}
$Mpc, and $M_{8} \propto \Omega$. 
For a CDM-type spectrum, 
n $ \approx -1 $ to $-2$  on scales of $\sim 8 \rm h^{-1}$ Mpc; 
for mixed (cold + hot) dark matter models, MDM, n $\sim -2.5$ (Klypin {\em et al.} 1993).

We calculate the cluster evolution by numerically
integrating eq.(1) (see below). First, however, we show how eq.(1) can be approximated,
in order to better understand the major factors that dominate the evolution.
For clusters of Richness $\gtrsim 1$ ($M > 3 \times 10^{14} \rm h^{-1}M_{\odot}$),
the exponential part of eq.(1)  
dominates the mass function.
Therefore,
\begin{eqnarray}
\frac{dn}{dM} &\propto &exp\left(\frac{-\delta_{c}^{2}}{2\sigma(z,M)^{2}}\right)
\end{eqnarray}
Transferring to an integrated mass function, we find approximately (c.f., Pen, 1997)
\begin{eqnarray}
\ln n(z,>M) & \propto & -(D(z,\Omega,\Lambda) \sigma_{0}(M))^{-2}  \nonumber \\
           & \propto & - \sigma_{8}^{-2} \Omega^{-2\alpha} D^{-2}M^{2\alpha}
\end{eqnarray}
We define the evolution rate ${\cal E}(z)$, for clusters with mass $> M$, as :
\begin{equation}
{\cal E}(z, >M)  =  -\frac{d\log_{10}n(z, >M)}{dz}  
\end{equation}
For small redshifts, the cluster abundance decreases approximately as $n(z) \propto 10^{-{\cal E}(z)(1+z)}$, with  $n(z)/n(0) = 10^{-{\cal E}(z)z}$
(see below). 
For masses of rich clusters, eq.(6) yields: 
\begin{equation}
{\cal E}(z, >M)  \propto  \sigma_{8}^{-2} \Omega^{-2\alpha} \frac{dD^{-2}(z,
\Omega,\Lambda)}{dz} M^{2\alpha}
\end{equation}
 
The evolution of the cluster abundance is therefore 
controlled by four  factors :

 1. $\sigma_{8}$;  
the evolution rate depends exponentially on $\sigma_{8}^{-2}$; 
clusters in  low $\sigma_{8}$ 
models evolve exponentially  faster than in  high $\sigma_{8}$ models (for a given $\Omega$). 
This arises since a fixed mass cluster corresponds to a higher 
relative density  peak (i.e., rarer fluctuation) in low $\sigma_{8}$ models.
This is the dominant factor in the  cluster evolution at recent times.

 2. $dD^{-2}/dz$; a linear perturbation grows slower in a low-density universe
than in a high-density one.
For $\Omega = 1$, the fluctuations grow as $(1+z)^{-1}$, 
thus continuing to evolve strongly at low $z$, while a low-density universe freezes out
fluctuations at an earlier epoch. 
High $\Omega$ models thus evolve faster than lower $\Omega$ models in recent times.
For the same $\Omega$, $dD^{-2}/dz$ is larger in a $\Lambda$-dominated
universe than in an open universe (Peebles 1980), therefore stronger evolution
is expected in a $\Lambda$-dominated model (see also \S 3).  

3. $\Omega^{-2\alpha}$; this factor arises because 
a fixed cluster mass $M$, 
corresponds to a different part of the power spectrum $P(k)$,
since $k^{-1} \sim R \sim (M/\Omega)^{1/3}$ is larger
for low $\Omega$ models. According to eq.(4),
$\sigma_{0}(M) \propto \Omega^{\alpha}$. 
For a given $\sigma_{8}$, a fixed mass cluster corresponds to a smaller 
$\sigma_{0}(M)$ for a lower $\Omega$, implying a rarer peak in the density field. 
Contrary to factor \#2 above, the contribution from
this term results in  faster evolution for low $\Omega$.
As we show below , the contributions from these two terms nearly cancel each other 
for $z \lesssim 1$.  
This cancellation can be also estimated directly from the approximate relations
at low $z$, $dD^{-2}/dz \propto f(\Omega) \simeq \Omega^{4/7} \simeq \Omega^{0.57}$
(Lightman \& Schechter 1990), and $\Omega^{-2\alpha} \propto \Omega^{-0.54}$ (for
$\alpha=0.27$, or $n=-1.4$), which approximately cancel each other.

4. $M^{2\alpha}$; this term is a constant for a given cluster mass.
The evolution is stronger for more massive clusters.

 We illustrate the above results in Figure 1, where  the
dependence of the evolution rate ${\cal E}(z)$ on $\Omega$ and $\sigma_{8}$ 
is calculated by numerically integrating eq.(1).
The results are  presented for a $\Lambda = 0$ CDM model at
redshifts 0 and 1 for clusters with virial mass $M \geq 3 \times 10^{14} \rm h^{-1}
M_{\odot}$ (corresponding to Abell richness class $\geq 1$, Bahcall \&
Cen 1993).
A spectrum with n=--1.4 is used, i.e. $\alpha=0.27$ 
(corresponding to a power spectrum shape parameter $\Gamma = \Omega h \simeq 0.25$, 
Eke {\em et al.} 1996,  Viana \& Liddle  1996).
 At $z \sim$ 0, the cluster evolution rate is
mainly determined by the amplitude $\sigma_{8}$ and is nearly independent of $\Omega$
 (or $\Gamma$, see below). 
For $\sigma_{8}=1$, ${\cal E}(z) \sim 1$, implying that 
the abundance of  richness $\geq 1$ clusters decreases by a factor of $\sim 3$ from
$z=0$ to $z=0.5$ ($\sim 10^{0.5{\cal E}(z)})$;
 for $\sigma_{8}=0.5$, 
${\cal E}(z) \sim 3.5$, and the 
number density decreases by a factor of $\sim 10^{2}$ for the same redshift interval.
${\cal E}(z)$ changes by only $\sim$ 10 \% when $\Omega$ changes
from 0.2 to 1; 
the contributions from the $dD^{-2}/dz$ and $\Omega^{-2\alpha}$ terms
described above (eq.8) approximately cancel
each other. The cluster evolution rate at $z < 1$,
${\cal E}(z) \propto \sigma_{8}^{-2}$ for fixed mass clusters,
can therefore be used to determine $\sigma_{8}$.
At $z \simeq$ 1, the evolution rate depends on both $\Omega$ and $\sigma_{8}$, but
$\sigma_{8}$ is still the dominant factor. 

Figure 2a shows the dependence of the evolution rate on the slope of the spectrum
n. For a CDM-type power spectrum, n $\sim -1$ to --2, the evolution is nearly 
independent of the shape of the spectrum (for any $\Omega$; see also White {\em et al.} 1993,
Eke {\em et al.} 1996, Pen 1997).
The evolution rate ${\cal E}(z)$ changes by $\lesssim 20 \%$ when n increases from --1 to --2. 
An MDM model (n $\sim -2.5$)  
evolves only slightly faster than Standard CDM models at this mass.
Figure 2b shows the
dependence of the evolution rate on  the cluster mass threshold.
From eq.(8) we see that  
${\cal E}(z) \propto  \sigma_{8}^{-2} M^{2\alpha}$.
The strong dependence of the cluster evolution on $\sigma_{8}$ and the milder dependence on
 mass threshold (for $\alpha \sim 0.3$) indicates that  
determining  $\sigma_{8}$ from cluster evolution  is not  
sensitive to reasonable uncertainties  in the cluster mass;
changing the cluster mass by a factor of two causes a $\lesssim 20 \%$
change in $\sigma_{8}$ (see Fig.3 below). 

Figure 3 presents the cluster abundance ratio $n(z=0)/n(z=0.5)$ as a function
of $\sigma_{8}$ for clusters with virial mass $M \geq 3 \times 10^{14} \rm h^{-1}
M_{\odot}$ (Richness $\gtrsim$ 1),
and $M \geq 6 \times 10^{14} \rm h^{-1} M_{\odot}$ (Richness $\gtrsim$ 2),
using an n=--1.4 power spectrum.
The dots represent all values of $\Omega$ from 0.2 to 1 for each $\sigma_{8}$.
The figure illustrates the strong dependence of the evolution rate on
$\sigma_{8}$, 
and the weak dependence on $\Omega$.
The relation can be approximated {\em for all $\Omega$ values}, as:
\begin{equation}
\frac{n(0)}{n(0.5)} =  
\left\{ \begin{array}{lll}
 	   10^{0.55/\sigma_{8}^2} \hspace{3mm}(\Lambda \neq 0), 
 	 & 10^{0.55\Omega^{0.15}/\sigma_{8}^2} \hspace{3mm}(\Lambda =0), 
	 & M \geq 3\times10^{14} \\
	   10^{0.8/\sigma_{8}^2} \hspace{5mm}(\Lambda \neq 0),     
	 & 10^{0.8\Omega^{0.15}/\sigma_{8}^2} \hspace{5mm}(\Lambda =0), 
         & M \geq 6\times10^{14}
 \end{array}
\right.
\end{equation}
These relations are represented by  the solid lines in Fig.3. 
More generally, the evolution of any rich cluster  
can be approximated as
\begin{equation}
\log_{10}n(0)/n(0.5) = 0.8 M^{2\alpha}_{6}/\sigma_{8}^{2}\hspace{3mm}
(\Lambda \neq 0),\hspace{5mm} 0.8\Omega^{0.15} M^{2\alpha}_{6}/\sigma_{8}^{2}\hspace{3mm}(\Lambda=0)
\end{equation}
where $M_{6}$ is the virial mass threshold in units of
$6\times10^{14}\rm h^{-1}M_{\odot}$, 
and $\alpha \simeq 0.27$ for CDM.
Observations of cluster abundance from $z=0$ to $z \sim$ 0.5 can thus 
determine $\sigma_{8}$ by either directly integrating eq.(1) (Fig.3--4)  or
by the approximate empirical relations (9--10).
\section{Determination of $\sigma_{8}$}
In a recent paper (Bahcall {\em et al.}  1997), we 
studied the evolution of the cluster mass function using  large scale N-body simulations.
Several cosmological models were studied : two Standard Cold Dark Matter models  (SCDM, $\Omega=1$,
$\sigma_{8} = $  1.05 and 0.53);
a low-density open CDM (OCDM, $\Omega$=0.35, $\sigma_{8}=0.79$);
a low-density $\Lambda$-dominated CDM(LCDM,  $\Omega$=0.40, $\sigma_{8}=0.80$),
and two  Mixed Dark Matter models (MDM, $\Omega=1$, $\sigma_{8}=0.60$ and $\sigma_{8}=0.67$). 
In Figure 4 we plot the relation between
the cluster abundance  ratio $n(z=0)/n(z=0.5)$ and $\sigma_{8}^{-2}$ as obtained from the N-body simulations, 
for clusters with mass $M_{1.5}\geq 6.3\times 10^{14}\rm h^{-1}M_{\odot}$
(within a physical radius of $1.5\rm h^{-1} Mpc$, as used in observations).
and compare it with the expected evolution (\S 2) of $\log n(z=0)/n(z=0.5) \propto \sigma_{8}^{-2}$.
The solid line is  calculated from the Press-Schechter relation (eq.1.) for
$\Lambda=0$ CDM model(for the mean of all $\Omega$'s, from 0.2 to 1),
for the same mass clusters (within physical radius of $1.5\rm h^{-1} Mpc$,
for proper comparison with the simulations and observations;
the comoving radius used in the previous paper is replaced here with
the  physical radius to better reflect the final observations). 
For the Press-Schechter approximation,
the mass is scaled from virial to $1.5\rm h^{-1} Mpc$  radius as follows:
the overdensity within the virial radius is numerically calculated for each model 
(e.g., Eke {\em et al.} 1996, Oukbir \& Blanchard 1997), 
thus yielding the virial radius for a given virial mass.
The virial mass in eq.(1) (i.e., mass within the virial radius) 
is then  scaled  to  the $1.5\rm h^{-1} Mpc$ mass using the
observed mass profile of $M \propto r^{0.64}$ (White {\em et al.} 1993,  
Carlberg {\em et al.} 1997).
The results are insensitive to this transformation since the two masses (radii) are comparable.
The dashed line in Fig.4 represents the best-fit to the model simulations; 
it agrees  well with the Press-Schechter relation.
Both Press-Schechter and the simulation show that 
the evolution rate is mostly dependent on $\sigma_{8}^{-2}$,
and is insensitive to $\Omega$ and  to the shape of the power spectrum. Figure 4 also shows that, as expected (\S 2), the LCDM model evolves somewhat faster
than OCDM, for have similar $\Omega$ and $\sigma_{8}$.  

The strong dependence of the cluster evolution rate on $\sigma_{8}$  
provides a powerful method for constraining $\sigma_{8}$
by comparison with the observed evolution of high-mass clusters to
$z \sim 0.5 - 1$. We  use the CNOC/EMSS sample of
high-redshift clusters (Carlberg {\em et al.} 1997, 
Luppino \& Gioia 1995) to determine the observed evolution
(see, e.g., Bahcall {\em et al.} 1997).
We find only mild evolution of cluster abundance. 
For massive clusters with $M_{1.5}\geq 6.3\times 10^{14}\rm h^{-1}M_{\odot}$ 
(within physical radius of $1.5\rm h^{-1} Mpc$) 
we find a mean best-fit observed $\log n(z) \propto z$ relation of
$\log n(z=0)/n(z=0.5) = 0.70 \pm 0.3 (1 \sigma)$. 
This observed evolution is compared with the ${\cal E}(z) - \sigma_{8}$
prediction in Fig.4 (shaded band).
The comparison yields a powerful
constraint on $\sigma_{8}$: 
$\sigma_{8} = 0.83 \pm 0.15$ (1 $\sigma$) 
(see also Bahcall {\em et al.} 1997).
These results are consistent with those obtained by Carlberg {\em et al.} (1997). 
Since the cluster evolution rate is not sensitive to the shape of the power spectrum,
the above determination applies to both CDM and MDM models.
Low $\sigma_{8}$ (highly biased) models  predict an extremely
low abundance of high mass clusters at $z \sim 0.5$, 
 inconsistent with the observed cluster abundance at $z \sim 0.5$.

The evolution of cluster abundance provides the first method that
     can directly determine the amplitude of the mass fluctuations
$\sigma_{8}$ independently of $\Omega$ or the power spectrum. Equivalently,
it provides the bias parameter on these scales.
The data  suggests that we live in a nearly unbiased
universe, with b $ \simeq \sigma_{8}^{-1} \simeq 1.2 \pm 0.2$,
and that mass approximately traces light on large scales.
The determination of  $\sigma_{8}$ 
breaks the degeneracy that exists between  $\sigma_{8}$ and $\Omega$  from the
present-day cluster abundance(\S 1 references; Bahcall {\em et al.} 1997).
Using the $\sigma_{8} - \Omega$ relation determined by Eke {\em et al.} (1996), 
and  the present constraint of $\sigma_{8} = 0.83 \pm 0.15$, 
we find $\Omega = 0.3 \pm 0.1$. 

We thank D. Eisenstein, B. Paczynski, D. Spergel and M. Strauss for helpful discussions.
This work is supported by NSF grant AST93-15368, ASC93-18185, and NASA grant
NAG5-2759. XF acknowledges support from an Advisory Council Scholarship.

\newpage
\begin{figure}
\vspace{-6cm}

\epsfysize=600pt \epsfbox{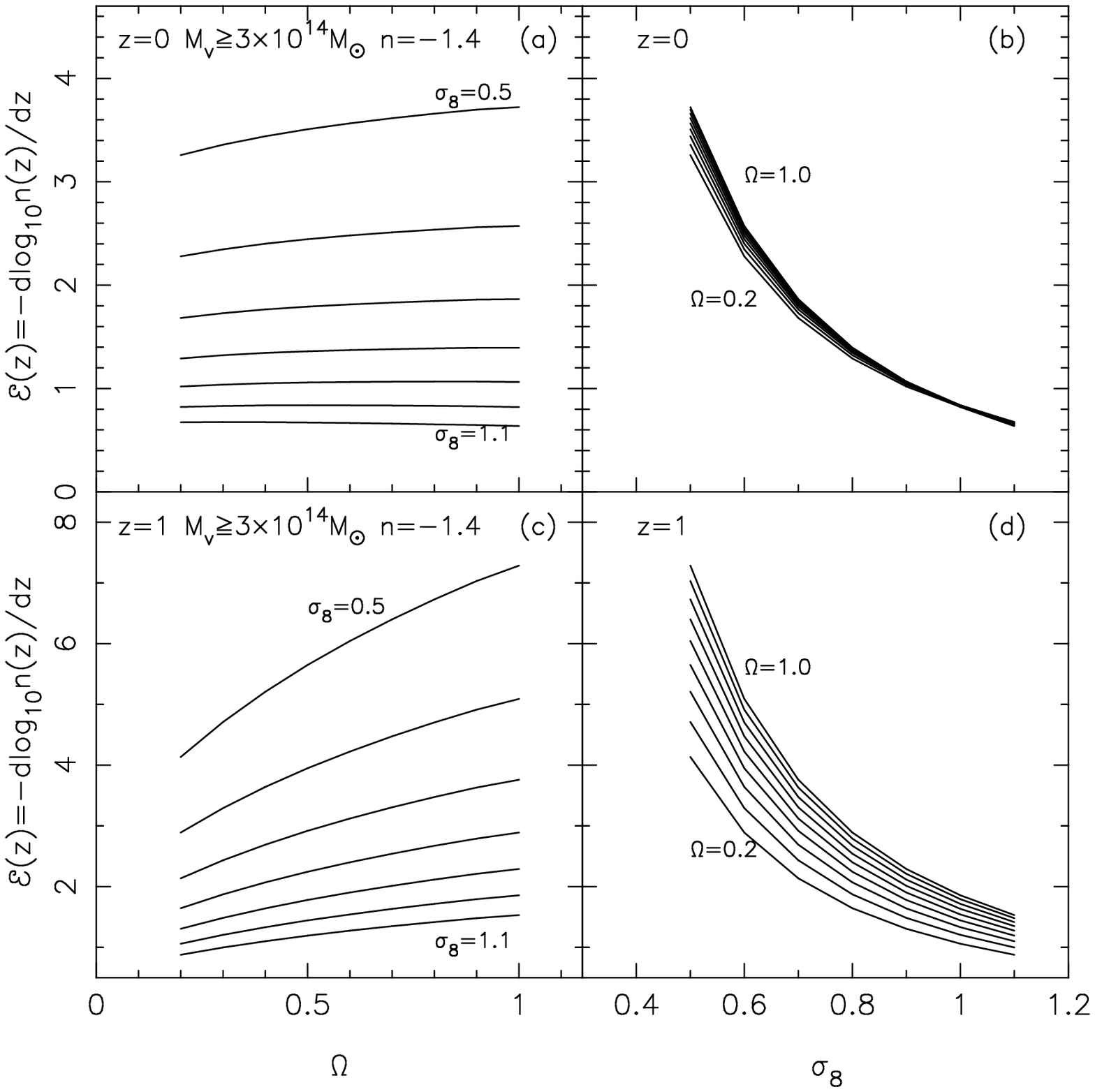}

\vspace{3cm}
Figure 1. Dependence of the evolution rate ${\cal E}(z)$ of
$M_{v} \geq 3 \times 10^{14} \rm h^{-1}M_{\odot}$ clusters
on $\Omega$ and $\sigma_{8}$ for $\Lambda=0$ models (at $z \sim 0$ and $z \sim 1$).
\end{figure}
\begin{figure}
\vspace{-6cm}

\epsfysize=600pt \epsfbox{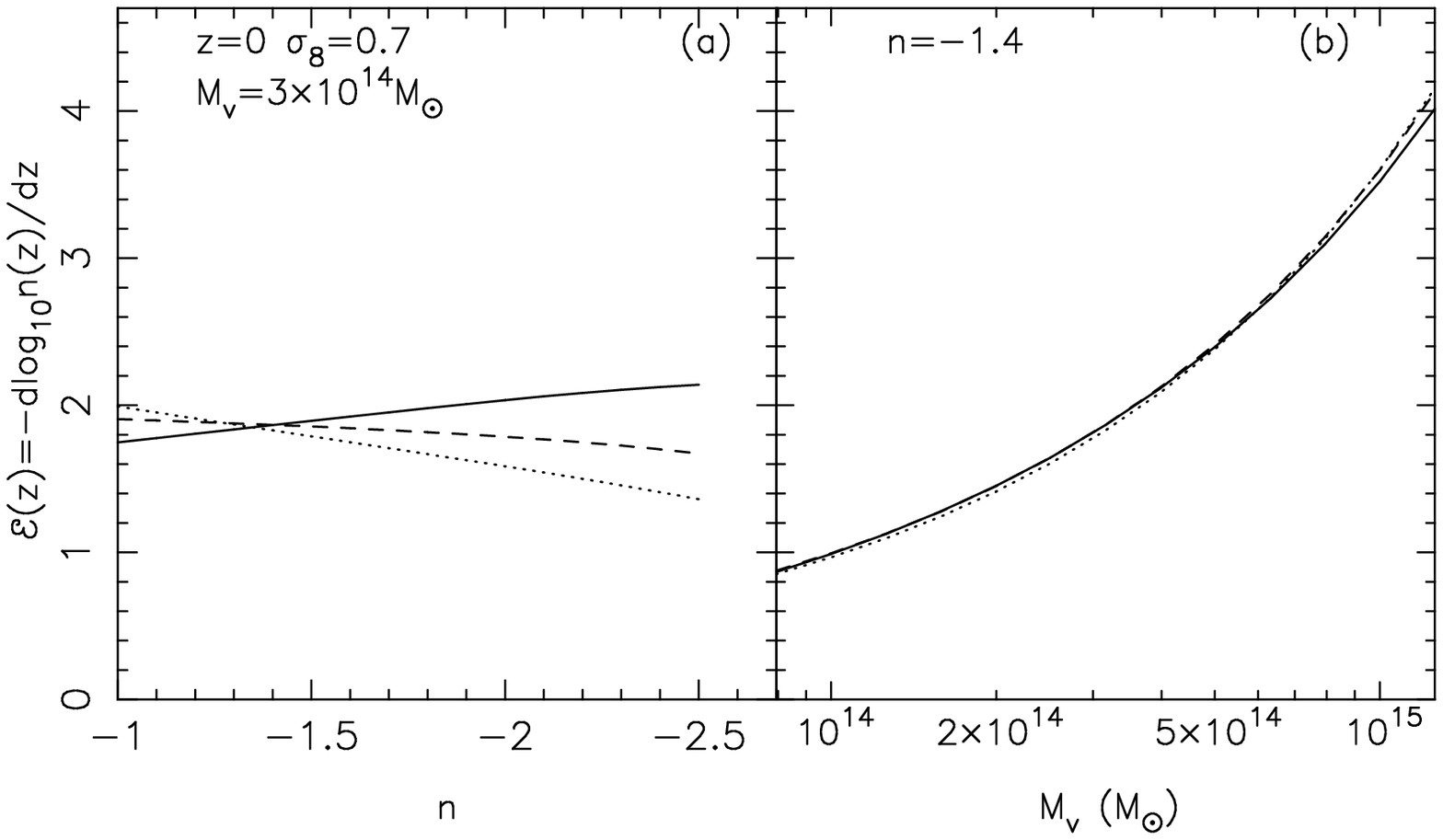}

\vspace{3cm}
Figure 2. Dependence of the evolution rate ${\cal E}(z)$ on the slope of the power
spectrum n and on the cluster mass threshold $M$.
Solid, dashed and dotted lines correspond to $\Omega=1$, 0.5 and 0.3 flat models,
respectively.
\end{figure}
\begin{figure}
\vspace{-6cm}

\epsfysize=600pt \epsfbox{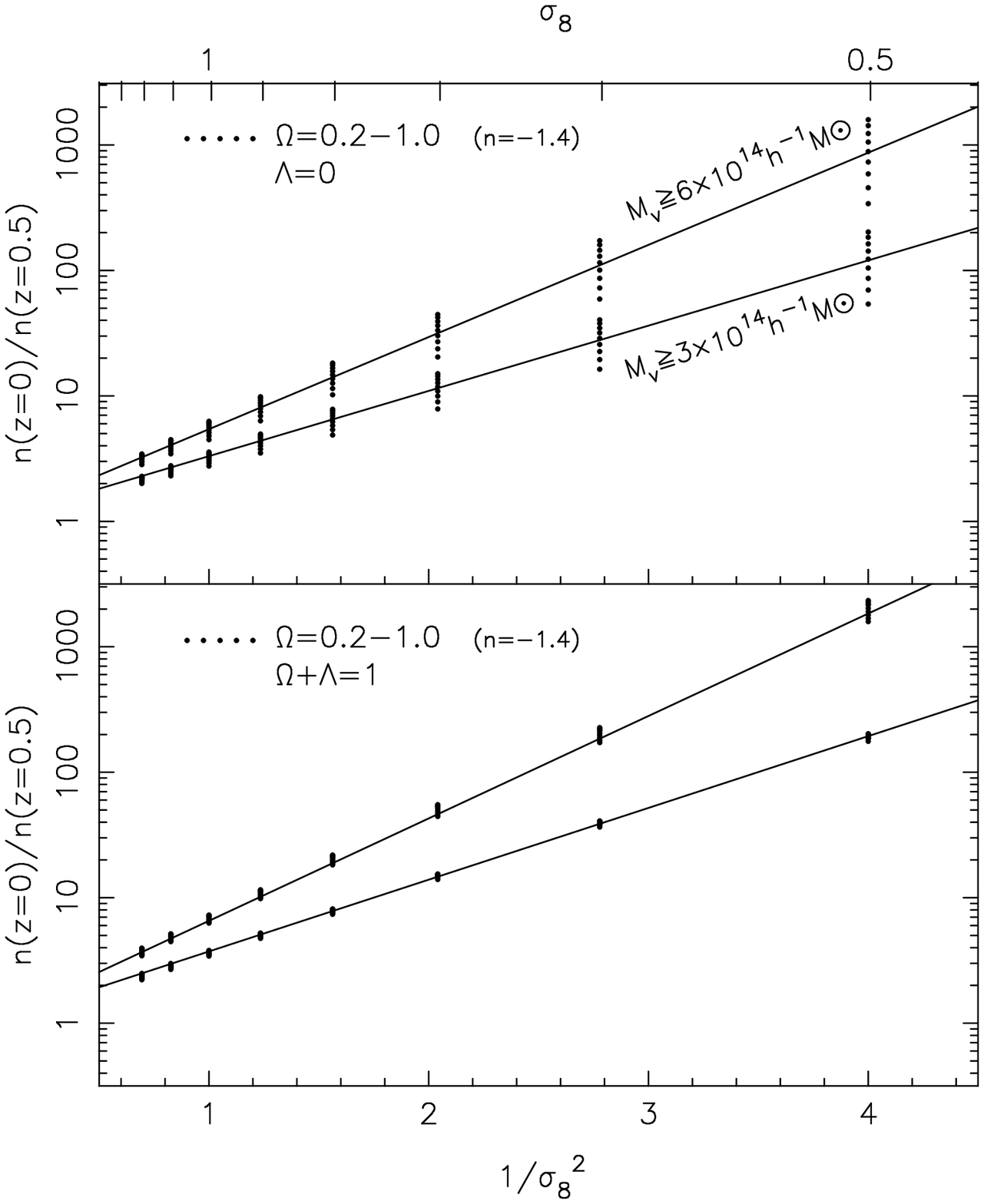}

\vspace{3cm}
Figure 3. Cluster abundance ratio $n(z=0)/n(z=0.5)$ versus
$\sigma_{8}$ for clusters with virial mass $M_{v} \geq 3 \times 10^{14} \rm h^{-1}
M_{\odot}$ (Richness $\gtrsim$ 1),
and $M_{v} \geq 6 \times 10^{14} \rm h^{-1} M_{\odot}$ (Richness $\gtrsim$ 2),
from the Press-Schechter relation eq.(1)
(represented by dots  for all values of $\Omega$ from 0.2 to 1 (bottom to top) for each $\sigma_{8}$).
The lines are the best-fit empirical relations (eq.9, with $\Omega \simeq 0.5$).
\end{figure}
\begin{figure}
\vspace{-6cm}

\epsfysize=600pt \epsfbox{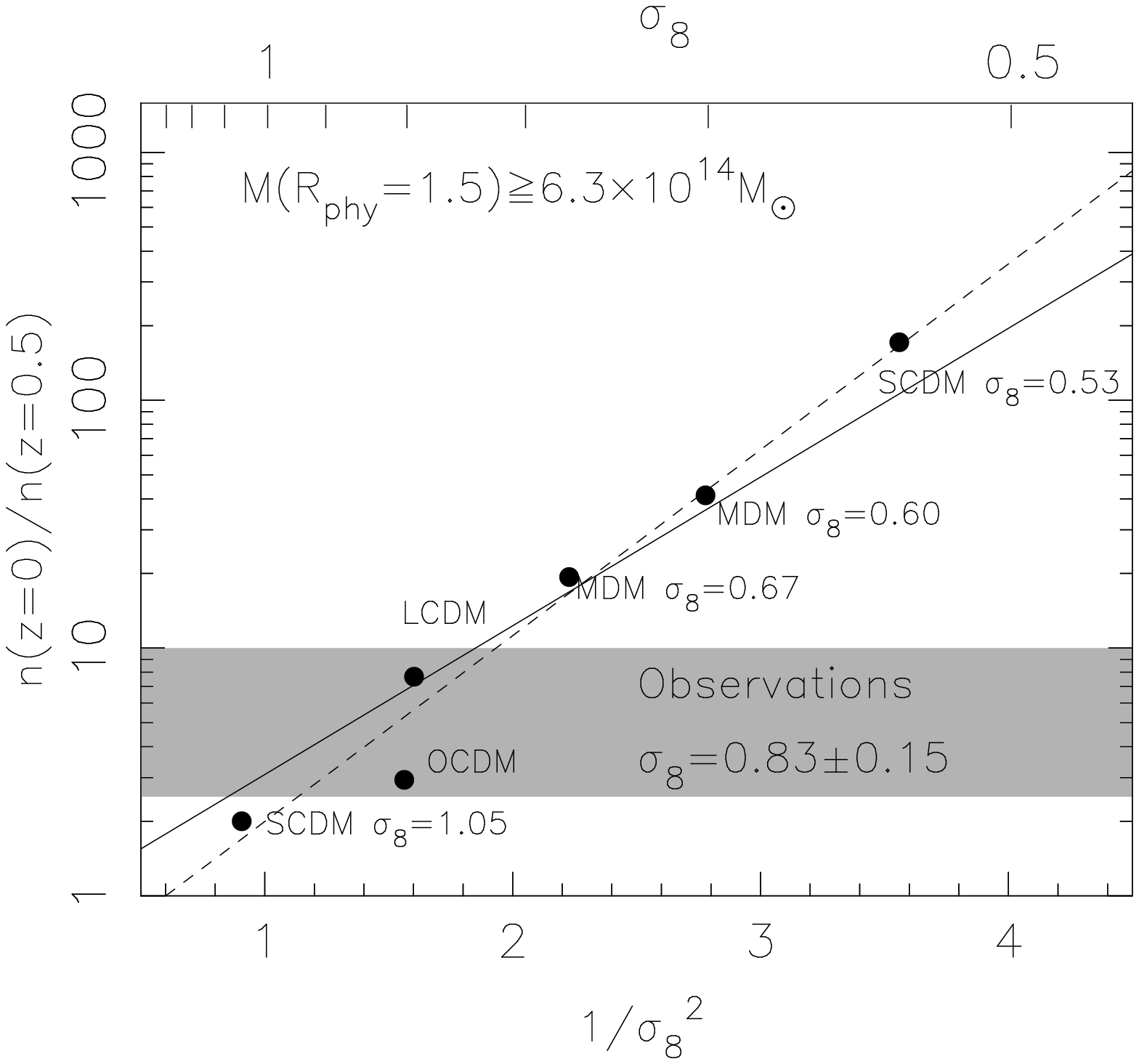}

\vspace{3cm}
Figure 4. Cluster abundance ratio $n(z=0)/n(z=0.5)$ versus
$\sigma_{8}$ from N-body simulations (dots),
and Press-Schechter approximation (solid line),
for $M_{1.5}\geq 6.3\times 10^{14}\rm h^{-1}M_{\odot}$  clusters.
The observed abundance ratio (\S 3) is shown by the shaded region.
(The solid line is the direct P-S relation (eq.1) for the given $M_{1.5}$ mass
threshold, for $\Lambda=0$ and mean of all $\Omega$'s;
the dashed line represents the best-fit relation from the simulations.)
\end{figure}
\end{document}